\documentclass[9pt,twocolumn,twoside]{pnas-new}

\templatetype{pnasresearcharticle} %
\setboolean{displaywatermark}{false}
\title{A unifying autocatalytic network-based framework for bacterial growth laws.}

\author[a,1]{Anjan Roy}
\author[a,2]{Dotan Goberman}
\author[a,b,3]{Rami Pugatch} 

\affil[a]{Department of Industrial Engineering and Management, Ben-Gurion University of the Negev,
Beer Sheva, 8410501, Israel}
\affil[b]{Quantitative Life Science Section, The Abdus Salam International Center for Theoretical Physics, Strada Costiera 11, 34014, Trieste, Italy}

\leadauthor{Roy} 

\significancestatement{Bacterial cells contain various autocatalytic cycles, such as the ribosome cycle, wherein ribosomes translate ribosomal proteins that subsequently self-assemble to form new ribosomes. Here, we show that the transcription--translation machinery couples all cellular autocatalytic cycles, resulting in balanced exponential growth. Each autocatalytic cycle gives rise to two types of growth laws. We derive the RNA polymerase growth law based on RNA polymerase autocatalytic cycle, whereas RNA polymerases transcribe mRNAs of Rpo subunits. Before degrading, these mRNAs catalyze Rpo proteins employing ribosomes. Then the Rpo proteins self-assemble to form new RNA polymerases, thus completing the cycle. We show that contrary to ribosome growth law, a reduction in growth rate due to shortage in RNA polymerase occurs without affecting the ribosomal protein mass fraction.}

\authorcontributions{Please provide details of author contributions here.}
\authordeclaration{Please declare any conflict of interest here.}
\correspondingauthor{\textsuperscript{3}To whom correspondence should be addressed. E-mail: rpugatch@bgu.ac.il}

\keywords{bacterial growth laws $|$ transcription $|$ translation $|$ autocatalysis $|$} 

\begin{abstract}
Recently discovered simple quantitative relations, known as bacterial growth laws, hint on the existence of simple underlying principles at the heart of bacterial growth. In this work, we provide a unifying picture on how these known relations, as well as new 
relations that we derive, stems from a universal autocatalytic network common to all bacteria, facilitating balanced exponential growth of individual cells. We show that the core of the cellular autocatalytic network is the transcription--translation machinery---in itself an autocatalytic network comprising several coupled autocatalytic cycles, including the ribosome, RNA polymerase, and tRNA charging cycles. We derive two types of growth laws per autocatalytic cycle, one relating growth rate to the relative fraction of the catalyst and its catalysis rate, and the other relating growth rate to all the time scales in the cycle. The structure of the autocatalytic network generates numerous regimes in state space, determined by the limiting components, while the number of growth laws can be much smaller. We also derive a growth law that accounts for the RNA polymerase autocatalytic cycle, which we use to explain how growth rate depends on the inducible expression of the rpoB and rpoC genes, which code for the RpoB and C protein subunits of RNA polymerase, and how the concentration of rifampicin, which targets RNA polymerase, affects growth rate without changing the RNA-to-protein ratio. We derive growth laws for tRNA synthesis and charging, and predict how growth rate depends on temperature, perturbation to ribosome assembly, and membrane synthesis.
\end{abstract}

\dates{This manuscript was compiled on \today}
\doi{\url{www.pnas.org/cgi/doi/10.1073/pnas.XXXXXXXXXX}}

\begin{document}

\maketitle
\thispagestyle{firststyle}
\ifthenelse{\boolean{shortarticle}}{\ifthenelse{\boolean{singlecolumn}}{\abscontentformatted}{\abscontent}}{}

\dropcap{T}he transcription--translation machinery is a universal set of molecular machines at the core of all known self-reproducing single-cell organisms. It can be considered as an embodiment of von Neuman's concept of a universal constructor---a machine capable of making other machines, self included, by reading an instruction set and consuming raw materials \cite{VN1, RP}. 

The transcription--translation machinery is composed of two key molecular machines, RNA polymerase and the ribosome. According to the central dogma, all cellular proteins are synthesized by this core machinery in a two-step process: RNA polymerases first transcribe genes to form mRNA ``instruction sets'', which are then translated by ribosomes to form proteins. 

To qualify as a ``universal constructor'', the transcription--translation machinery should also be capable of replicating itself. The self-replication of the transcription--translation machinery is a complex process, which is, nevertheless, universal to all single-cell organisms capable of self-replication. It proceeds via two prominent coupled autocatalytic cycles, the RNA polymerase autocatalytic cycle, and the ribosome autocatalytic cycle. The two cycles are coupled because the de-novo synthesis of new ribosomes cannot take place without RNA polymerase transcribing ribosomal RNA (rRNA), while the de-novo synthesis of RNA polymerase can not take place without ribosomes translating the mRNAs of rpo genes, which code for the RNA polymerase protein subunits that comprise the RNA polymerase. Each of these two cycles also involves a self-assembly step. 

Both the RNA polymerase and the ribosome autocatalytic cycles also rely on other autocatalytic cycles which are integral parts of the transcription--translation machinery. These additional autocatalytic cycles are responsible for charging tRNA with amino acids (aa) and in assisting the ribosomes to initiate, to translocate, and to terminate the translation process. The autocatalytic nature of these cycles is less familiar and becomes more evident when we consider each of its elements, e.g., tRNA as catalyzing itself with the help of RNA polymerases and ribosomes, as we explain below. 

All the autocatalytic cycles mentioned above are intertwined and require each other to perform autocatalysis; removing any key catalyst from any one of these cycles breaks autocatalysis in all the cycles.
\begin{figure}[ht]
\centering
\includegraphics[width=1\linewidth]{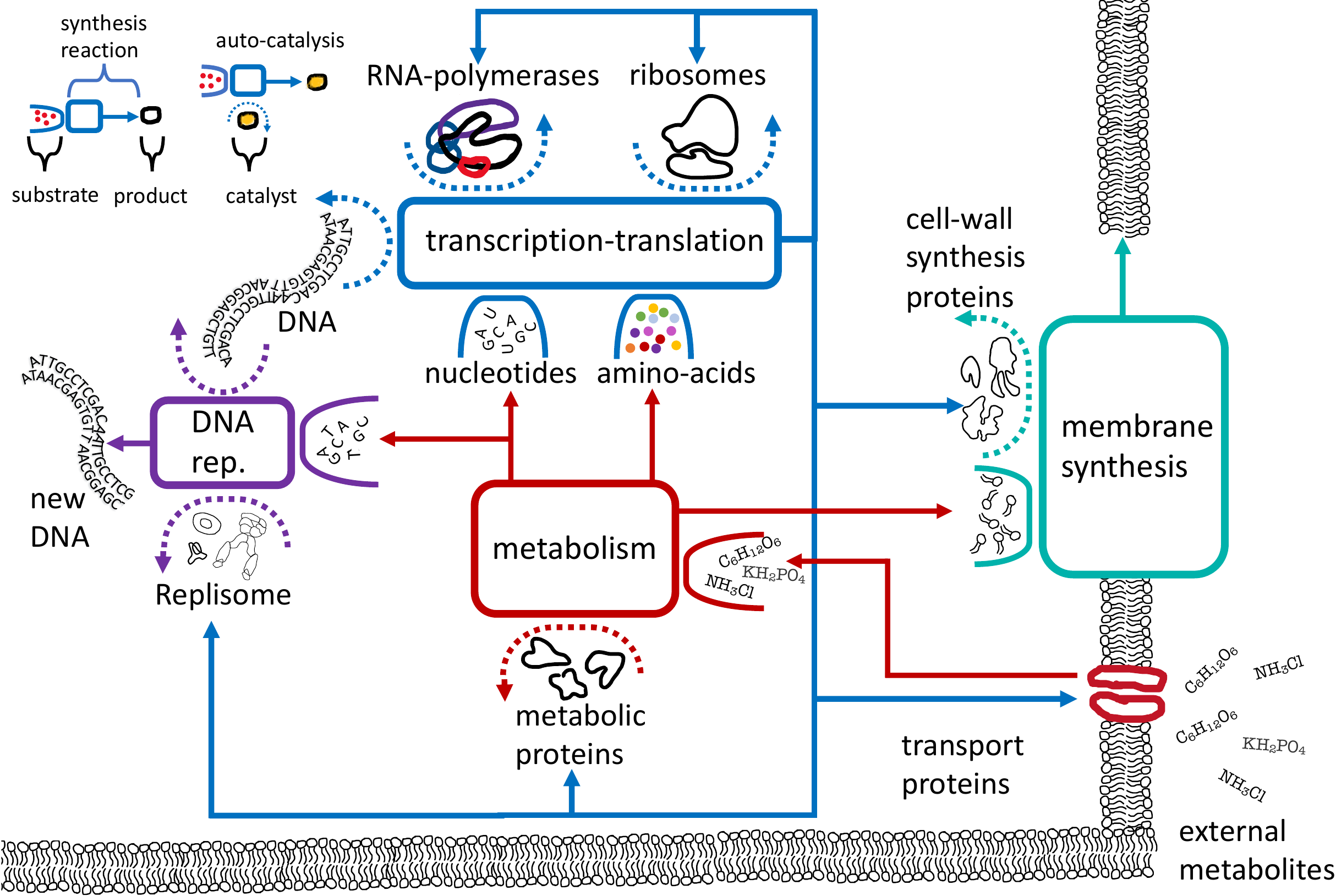}
\caption{Schematic diagram of a bacterial autocatalytic network, showcasing different autocatalytic cycles coarsely grained. Top left corner: An explanation of the graphical notation, following \cite{Hordijk}. A reaction node is marked by a square. Substrates consumed by the reaction are depicted inside open boxes. Catalysts that drive a reaction but are not consumed by it are depicted inside dashed curved arrows. Each arrow emanating from a reaction node points to a products of the synthesis reaction. In an autocatalytic reaction, the synthesized products themselves serve as the catalysts that drives the reaction.  In the Methods section, we explain how this graphical notation is translated to a set of coupled ordinary differential equations, from which we derive the growth laws by solving for the steady growth condition. In the main figure, we present a schematic autocatalytic reaction network for an entire cell. The transcription--translation machinery consumes raw materials and energy (not shown) and produces copies of all the proteins, including copies of itself. DNA is replicated by the replisome machinery, using the existing DNA as a template . Metabolic proteins import and convert external metabolites into nucleotides, amino acids, fatty acids, and other metabolites. The membrane is synthesized by the membrane synthesis proteins. Thus, four major autocatalytic cycles are shown in this coarse-grained picture---the autocatalysis of the transcription--translation machinery, of the DNA, of metabolism, and of the membrane. All these autocatalytic cycles are coupled by the transcription--translation machinery.} 
\label{fig1}
\end{figure}

Recently, the ``ribo-centric'' view, which focuses on the ribosome autocatalytic cycle (``ribosomes make ribosomes''), has led to the discovery of a bacterial growth law that quantitatively relates bacterial growth rate to the ribosomal protein fraction and the ribosome translation rate \cite{hwa1,RNA2prot}. A recent study focused on the relationship between growth rate, translation rate, and the transcription of rRNA \cite{Shlomi}. 

Despite its successes, the ``ribo-centric'' approach also has shortcomings, as it disregards both transcription---an important pillar in the central dogma---and other autocatalytic cycles in the cell. In certain cases discussed below, e.g., when the temperature mildly changed or when transcription is perturbed, a significant change is observed in the growth rate, but this change is not accompanied by a change in the ribosome fraction, as expected from the ribosome growth law presented in \cite{hwa1}. Explaining this deviation requires a more general approach.

Here, we take such a general approach by considering both transcription and translation on an equal footing, and we derive growth laws that are based on the autocatalysis of the transcription--translation machinery. Furthermore, we show that both transcription and translation couples all other autocatalytic cycles in the bacterial cell, leading to balanced exponential growth of all components, at the same growth rate, without requiring complex feedback mechanisms (see Fig. \ref{fig1}). 

We demonstrate that each autocatalytic cycle leads to two types of growth laws. The first type, to which we refer as the relative abundance growth law, relates the growth rate to the relative abundances of the catalysts that drive the cycle and to their catalysis rates. The second type, to which we refer as closed-cycle growth laws, relate the growth rate to all the catalysis rates and allocation parameters within a given autocatalytic cycle. An allocation parameter is the fraction of catalysts allocated towards a particular task, e.g., the fraction of ribosomes allocated to make ribosomal proteins. Using this formalism, we derive and discuss new growth laws, and we also re-derive existing growth laws, thus demonstrating the merits of a holistic picture of bacterial cellular growth. We show that the universal coupling induced by the transcription--translation machinery is responsible for locking all cycles to the same exponential growth rate, irrespective of the nature of the coupling which can be non-optimal (see Methods). 

The fact that global coupling, by itself, can guarantee balanced growth with a common growth rate implies that the biological function of well known feedback mechanisms, such as the stringent response \cite{Sauer,Stringent} or product feedback inhibition in metabolism \cite{FeedbackInhibition} and in ribosome assembly \cite{Nomura1}, are required for optimizing the coupling, e.g., for growth rate or efficiency, rather than for growth coordination.  

Our modeling approach offers a simple way to recognize ``limitation regimes'', characterized by a complete list of catalysts and substrates that locally limit the reactions in which they participate. The number of limitation regimes combinatorially explodes with the number of reactions accounted for by the model. Many limitation regimes can be further aggregated to form ``growth regimes'', which are characterized by having a common limiting autocatalytic cycle. As mentioned above, each limiting cycle gives rise to two types of growth laws. 

Notably, despite their elegant simplicity and experimental success \cite{hwa1,RNA2prot,ROPP}, bacterial growth laws do not uniquely define the cellular state, nor elucidate the complex evolutionarily shaped control mechanisms that drove the cell to a particular limitation and growth regime. Finding such an evolutionary design-logic remains an interesting open challenge \cite{ROPP}.

\section*{Results}

\subsection*{The transcription--translation machinery self-replicates using several coupled autocatalytic cycles}
\begin{figure}[h]
\centering
\includegraphics[width=1\linewidth]{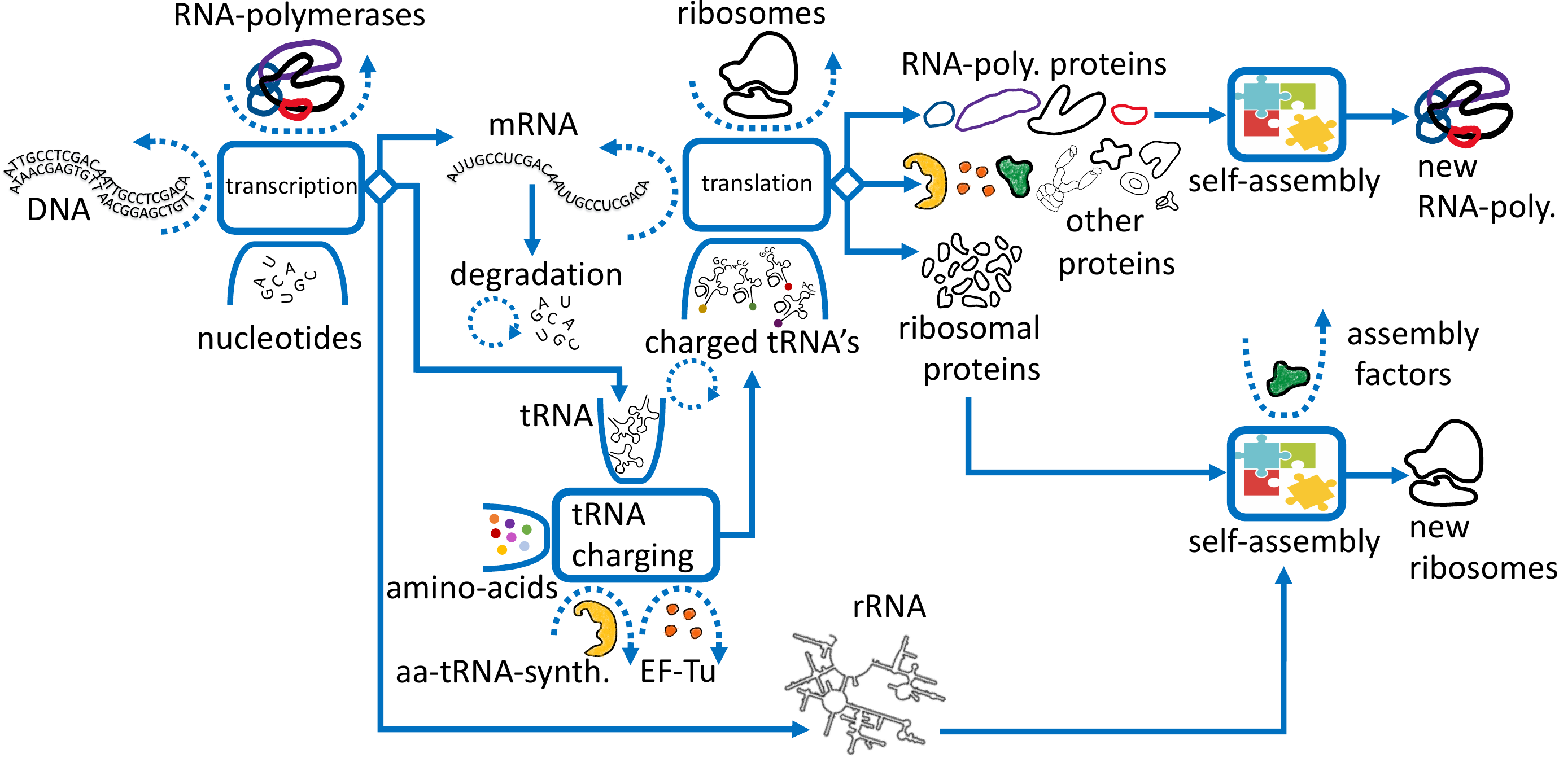}
\caption{The transcription--translation autocatalytic network. In this schematic diagram, we coarsely show how the transcription translation machinery self-replicates via three main coupled autocatalytic cycles: (a) the ribosome autocatalytic cycle, which comprises two reactions: one for ribosomes that synthesize ribosomal proteins and one for RNA polymerase that synthesizes rRNA. The ribosomal proteins and the rRNAs merge in a self-assembly reaction to form new ribosomes; (b) the RNA polymerase autocatalytic cycle, in which RNA polymerases transcribe the mRNAs that catalyze the production of the Rpo protein subunits, which, in turn, self-assemble to form new RNA polymerases; and (c) the tRNA-charging reaction where, e.g., aminoacyl-tRNA synthetases (aa-tRNA-synt.) catalyzes the charging of tRNA with amino acids, which, in turn, are transferring the amino acids to ribosomes that translate mRNAs, including the mRNAs of aminoacyl-tRNA synthetases. Any substrate that is not consumed by the reaction can be considered as a catalysts; for example, mRNA can be viewed as a catalyst for protein synthesis, but its catalysis rate is $n_i$ times higher than that of a single ribosome, where $n_i$ is the average number of ribosomes co-translating this mRNA. Importantly, in the absence of any type of material inputs---either catalysts like tRNAs, mRNAs, ribosomes, RNA polymerases, aminoacyl-tRNA synthetases, or substrates like amino-acids or rRNAs---will bring the autocatalysis of the entire network to a halt.}
\label{fig2}
\end{figure}

The transcription--translation machinery self-replicate using three main coupled autocatalytic cycles: (i) the ribosome cycle; (ii) the RNA polymerase cycle; and (iii) the tRNA-charging cycle; other translation--facilitating cycles are not discussed here. In Fig.
\ref{fig2} these three autocatalytic cycles are schematically depicted and explained.  

\textbf{The ribosome autocatalytic cycle.}
To synthesize ribosomes de-novo, existing ribosomes must create more than $50$ different ribosomal protein subunits. The mRNAs for the ribosomal proteins are transcribed by RNA polymerases. A sub-group of the ribosomal proteins directly bind to ribosomal RNA (rRNA), which is also transcribed by RNA polymerases. Subsequently, other ribosomal proteins bind to the sub-assembled ribosome, in a predefined partial order \cite{poset}, elucidated by the well known small and large ribosome subunit assembly maps \cite{Nierhaus, WilliamsonReview}.  

We derive the autocatalytic cycle of ribosomal proteins by assuming that rRNA is abundant and by focusing on the fraction $\alpha_{RP}$ of ribosomes that are allocated to synthesize these proteins. The ribosomal proteins spend some time free-floating and eventually enter the ribosome assembly line, where they spend some time in the assembly process and then exit, embedded in the small or large subunit of the ribosome. The newly synthesized small and large ribosome subunits spend some time free-floating before binding to mRNA and becoming engaged in the translation process.

We derive both the abundance and the closed-cycle growth laws by writing a set of coupled ordinary differential equations (ODEs) for the rate of change in the abundance of ribosomal proteins, the rate of conversion to an active assembling state, the rate of assembly of new ribosomes, and the rate of interconversion between resting and active states. The resulting closed-cycle growth law is
\begin{align}
(\mu \tau_{SA(R)} + 1) (\mu \tau_{pool(RP_j)} + 1)\left(\mu  + \frac{1}{\tau_{life(R)}}\right)  = \frac{\alpha_{RP_j}}{\tau_{RP_j}}\frac{R_{b}}{R} \label{eqRpACtot}
\end{align}
where $\mu$ is the growth rate, $\tau_{SA(R)}$ is the ribosome assembly time, $\tau_{pool(RP_j)}$ is the duration that the ribosomal protein spend in its assembly precursor pool, $R$ is the total number of ribosomes, $R_b$ is the total number of active ribosomes, $\tau_{life(R)}$ is the lifetime of the ribosome, $\alpha_{RP_j}$ is the fraction of ribosomes that are allocated to synthesize ribosomal protein $RP_j$, $\tau_{RP_j}=L_{RP_j}\tau_{aa}$ is the translation duration of ribosomal protein $RP_j$, whose lengths is $L_{RP_j}$ amino acids, and $\tau_{aa}^{-1}$ is the ribosome elongation rate.

If we further assume that both the ribosome assembly time and the duration that free-floating ribosomal proteins spend in their precursor pools are negligible compared with the doubling time, i.e., $\mu \ll 1/\tau_{SA(R)}$ and  $\mu \ll 1/\tau_{pool(RP_j)}$, we obtain
\begin{align}
\mu \tau_{RP_j} + \frac{\tau_{RP_j}}{\tau_{life(R)}}  = \alpha_{RP_j} \phi_b,
\label{eq2}
\end{align}
where $\phi_b=\frac{R_{b}}{R}$ is the fraction of active ribosomes . 

The term $\alpha_{RP_j} \phi_b$ in Eq. \ref{eq2} is the fraction of active ribosomes that are allocated to translate ribosomal proteins. Thus, the second term on the left-hand side stands for the allocation of active ribosomes to the translation of ribosomal proteins at zero growth $\mu=0$.

Equation \ref{eq2} is thus equivalent to the well known bacterial growth law, $\frac{\mu}{\gamma}  + \phi_0 = \phi_R$, presented and experimentally tested in \cite{hwa1,RNA2prot} under the conditions $\mu \tau_{SA(R)} \ll 1$ and $\mu \tau_{pool_i} \ll 1$. In this approximation, the mass fraction of the ribosomal proteins (which is $2/3$ of the RNA-to-protein ratio) is equivalent to the allocation parameter of the ribosomes, namely, the fraction of ribosomes allocated to translate ribosomal proteins (SI). Significantly, the validity of these assumptions, and the agreement between ribosome allocation and ribosome mass fraction, can be tested experimentally because ribosome allocation can be measured directly through ribosome profiling experiments, while ribosomal protein mass fraction can be measured using RNA-seq combined with mass-spectrometry \cite{RNA2prot}. Indeed, in a $20$ minutes doubling time, the measured ribosomal mass fraction was measured to be $30\%$ \cite{hwa1}, while a ribosome profiling experiment found that $28.5\%$ of the active ribosomes, or $\sim 32\%$ of the total number of ribosomes, are engaged in the process of translating ribosomal proteins \cite{RibosomeProfiling2}. Nevertheless, deviations from the correspondence between mass fraction and the allocation parameter are predicted by our model when one or more of the aforementioned assumptions breaks down, e.g., when the ribosome assembly time increases, as in \cite{Williamson1}. These predicted deviations can be experimentally tested. 

Since ribosomes are synthesized by the self-assembly of rRNA and ribosomal proteins, we can also write a ribosome growth law that is based on the production of rRNAs by RNA polymerases. The resulting abundance and closed-cycle growth laws are obtained by writing a set of coupled ODEs, for the rate of change in the abundance of RNA polymerase subunits, the rate of conversion to an actively transcribing state, the rate of transcription of new ribosomal RNA by RNA polymerases, and the rate of production of new ribosomes by these ribosomal RNAs after assembly (SI). We obtain the following closed-cycle growth law,  
\begin{align}
(\mu \tau_{SA(Rpol )} + 1) (\mu \tau_{SA(R)} + 1) (\mu \tau_{pool(Rpo_j)} + 1) \nonumber \\
\left(\mu R  + \frac{R}{\tau_{life(R)}}\right) \left(\mu Rpol  + \frac{Rpol}{\tau_{life(Rpol)}}\right)  \nonumber\\
= \frac{\alpha_{rRNA_j}  \alpha_{Rpo_j}}{\tau_{rRNA_j} \tau_{Rpo_j}}R_{b}Rpol _{b},
\end{align}
where $\tau_{rRNA_j}$ is the transcription time of the $j^{th}$ ribosomal RNA, $ \tau_{Rpo_j}$ is the translation time of the $j^{th}$ RNA polymerase subunit, $\tau_{SA(Rpol)}$ and $\tau_{life(Rpol)}$ are the assembly time and lifetime or the RNA polymerase, respectively, $\alpha_{Rpo_j}$ is the fraction of active ribosomes translating the $Rpo_j$ protein subunit of the RNA polymerase, $\alpha_{rRNA_j}$ is the fraction of active RNA polymerases transcribing $rRNA_j$, and $Rpol_b$ is the number of RNA polymerases that are actively transcribing.

If we assume further that  $\mu \ll 1/\tau_{SA(R)}$, $\mu \ll 1/\tau_{SA(Rpol )}$, $\mu \ll 1/\tau_{pool(Rpoj)}$, and that the lifetimes of ribosomes and RNA polymerases are longer than the doubling time, we obtain
\begin{align}
\mu^2 = \frac{\alpha_{rRNA_j}  \alpha_{Rpo_j} }{\tau_{rRNA_j} \tau_{Rpo_j} }\frac{R_{b}}{R}\frac{Rpol _{b}}{Rpol}
\end{align}
which is equivalent to the ribosome RNA autocatalysis growth law presented in \cite{Shlomi}. 

\textbf{The RNA polymerase autocatalytic cycle.}
\begin{figure}
\centering
\includegraphics[width=1\linewidth]{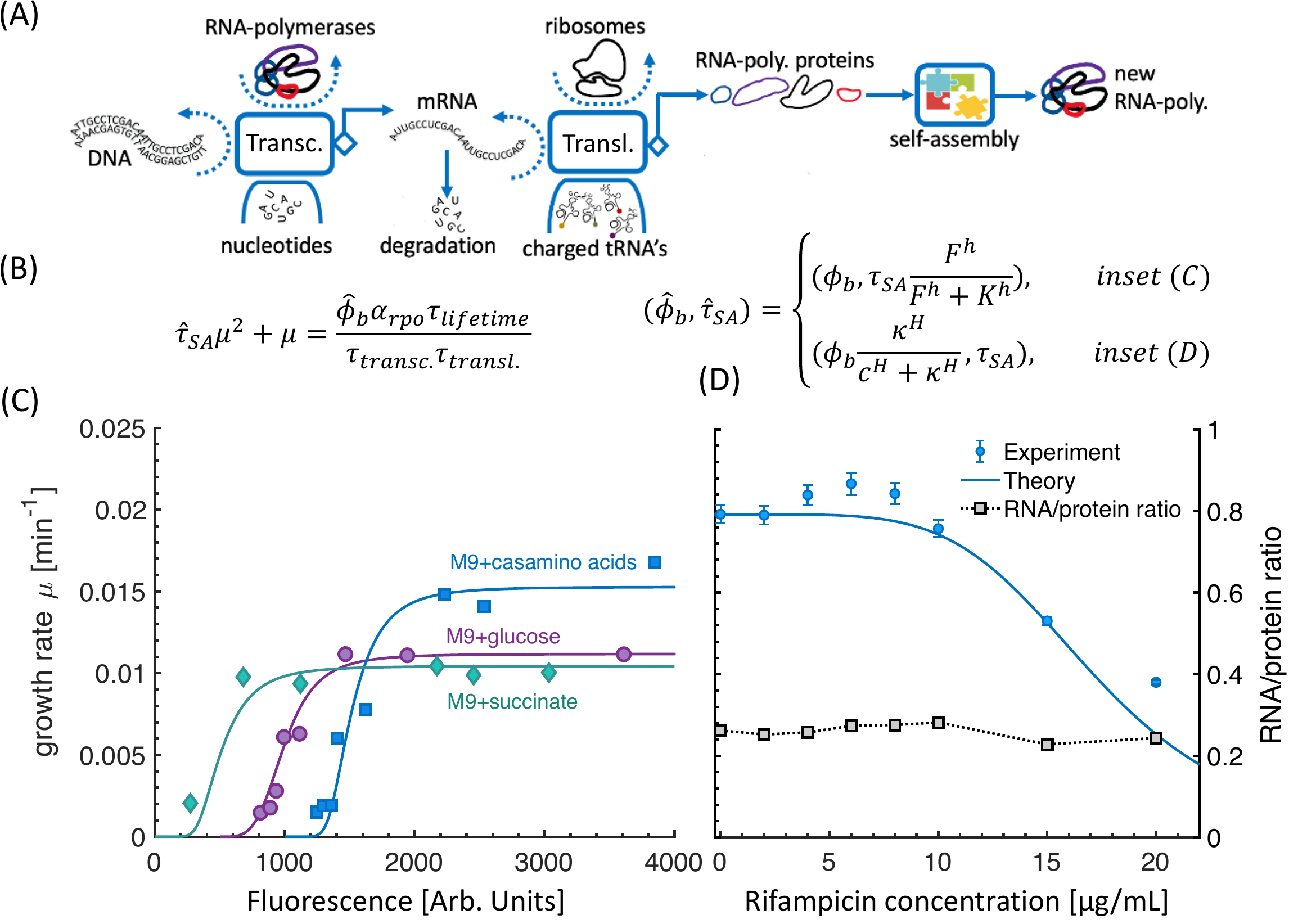}
\caption{The RNA polymerase growth law. (A) Illustration of the RNA polymerase autocatalytic cycle. Among the active RNA polymerases, a fraction $\alpha_{rpoA-D / Z}$ is allocated to transcribe the rpoA-D and rpoZ genes (rpoD not shown). The transcribed mRNAs produce the RpoA-D and RpoZ protein subunits during their lifetime. The rate of protein synthesis by mRNA is equal to $R_{m_i} \times \tau_{transl.}^{-1}$, where $R_{m_i}$ is the average number of ribosomes on an mRNA of type $i$ and $\tau_{transl.}=L_i \tau_{aa}$ is the duration for a single ribosome to translate mRNA of type $i$, whose length is $L_i$ base-pair triplets, and $\tau_{aa}^{-1}$ is the ribosome elongation rate. The $Rpo_j$ proteins, $j\in\{A,B,C,D,Z\}$ self-assemble to form new RNA polymerases, which join the collective pool of RNA polymerases. The fraction of active RNA polymerases $\phi_b$ is taken from \cite{Bremer, bionumbers}. (B) The RNA polymerase growth law. $\tau_{lifetime}$ is the rpoB mRNA life time, taken to be $3$ minutes \cite{taniguchi}, $\tau_{transc.}$ is the duration of transcription, calculated based on the transcription rate from \cite{bionumbers} and the length of the rpoB gene. The fraction of active RNA polymerases and the RNA polymerase assembly duration are modulated using separate Hill functions, in accordance with the experiment under consideration (see C and D for details). (C) The growth rate as a function of the fluorescent reporter protein, induced subsequently to the induction of the rpoB and rpoC genes in the experiment detailed in \cite{RpoSwitch}. Since RpoB precedes RpoC in the assembly of RNA polymerase, upon the depletion of the RpoB pool, the synthesis rate of RpoB governs the RNA polymerase assembly duration. The theoretical fits were produced using the RNA polymerase growth law (see B), by fitting only once the $K$ and $h$ parameters (see curly brackets in B). To facilitate the comparison between the theoretical fits and the data, we artificially shifted the M9+glucose by $500$ arbitrary units of fluorescence and the M9+casamino acids by $1000$ arbitrary units of fluorescence. (D) The effect of rifampicin on the growth rate of \textit{E. coli}; data taken from \cite{Jun17}. As rifampicin levels increase, the fraction of actively translating RNA polymerases, $\phi_b$, decreases. We find that, when the concentration of rifampicin $c$ is $c \approx 17$ $\mu g/mL$, the growth rate is reduced by half compared with the nominal ($c=0$) case. The measured RNA-to-protein ratio (which is proportional to the ribosomal protein mass fraction) remains constant, indicating that ribosomes do not limit the growth rate in this experiment. This is because RNA transcription becomes limiting, which equally attenuates both ribogenesis and protein synthesis due to a global shortage in all forms of mRNA, as explained in the main text.}
\label{Rpol fig}
\end{figure}

In bacteria, RNA polymerase comprises four core protein subunits --- rpoA ($\alpha$), rpoB ($\beta$), rpoC ($\beta'$), and rpoZ  ($\omega$) and an interchangeable $\sigma$-factor. Consider the ``RNA polymerase makes RNA polymerase'' autocatalytic cycle, which operates as follows. A fraction of the active RNA polymerases, $\alpha_{rpo}$, is allocated to transcribing mRNAs from the rpo genes, while a fraction of the active ribosomes, $\alpha_{Rpo}$, is allocated to translating these mRNAs and synthesizing the RNA polymerase protein subunits. Translation from a specific mRNA continues until the mRNA degrades, as it has a finite lifetime of $\sim 3$ minutes \cite{taniguchi}. The Rpo protein subunits subsequently self-assemble to form new RNA polymerases. 

The closed-cycle RNA polymerase growth law is obtained by writing a set of coupled ODEs for the rate of change in the mRNAs transcribed from the rpo genes, while accounting for their finite lifetimes, the rate of change in Rpo protein subunits, the rate of production of new RNA polymerases from the Rpo subunits after assembly, and the fraction of active RNA polymerases that are (re-)allocated to transcribe the rpo genes, thus sustaining the exponential growth of RNA polymerases in the cell. These ODEs yields the following closed-cycle growth law,
\begin{align}
\Pi_{i=1}^{4}(1+\mu \tau_i) = \frac{ \tau_{life(Rpol)} \alpha_{rpo_j} R_{m(Rpo_j)}  \tau_{life(m(Rpoj))} }{\tau_{Rpo_j} \tau_{rpo_j} } \tilde{\phi}_b
\label{RNAPgl}
\end{align}
where $\tau_1=\tau_{SA(Rpol)}$ is the RNA polymerase assembly duration, $\tau_2=\tau_{life(m(Rpo_j))}$ is the lifetime of the mRNA of $Rpo_j$, $\tau_3=\tau_{pool(Rpo_j)}$ is the duration that $Rpo_j$ spends in its precursor pool, and $\tau_{life(Rpol)}$ is the RNA polymerase lifetime. Additionally, $\alpha_{rpo_j}$ is the fraction of active RNA polymerases transcribing the mRNAs of the RNA polymerase subunit $Rpo_j$, $R_{m}$ is the average number of ribosomes translating this mRNA, and $\tilde{\phi}_b=\frac{Rpol_{b}}{Rpol}$ is the fraction of active RNA polymerases. As the growth rate decreases toward zero, Eq. \ref{RNAPgl} predicts that the cell will still contain a finite fraction of active RNA polymerases, as was the case in the ribosome growth law. 

To demonstrate the merit of our RNA polymerase growth law, we used data from two recent experiments \cite{Jun17, RpoSwitch}. In the first experiment, the researchers developed a reversible growth switch in \textit{E. coli} by removing rpoBC genes from its genome and placing them on a plasmid with an inducible promoter and a fluorescent reporter \cite{RpoSwitch}. The expression of rpoBC genes was thus controlled via the external concentration of the inducer, IPTG. When the IPTG concentration was high, nominal growth rates were observed. However, when the IPTG concentration dropped, a rapid decrease in growth rate was observed.

To explain the rapid decrease in growth rate, we employ our RNA polymerase growth law, Eq. \ref{RNAPgl}, with RpoB as the limiting factor in the assembly process. As long as the levels of RpoB under steady growth conditions are non-zero, the growth rate does not change. However, as soon as the free RpoB pool vanishes due to the induced reduction in its expression, the assembly duration of new RNA polymerases starts to increase. To account for this scenario, we assume that the increase in the assembly time equals to the delay in the delivery of RpoB. 

The measured fluorescence of the reporter protein is not proportional to the size of the free Rpo pool, because the RpoB proteins are consumed by the assembly reaction at a faster rate than the decay rate of the fluorescence. Therefore, we expect the fluorescence level to monotonically increase with the expression level. The equation that we use for the assembly duration is $\hat{\tau}_{SA(Rpol)}= \tau_{SA(Rpol)} \frac{F^h}{F^h+K^h}$, where $\tau_{SA(Rpol)}$ is the assembly duration in the nominal (high IPTG or wild-type) case. 
\begin{figure}[t]
\centering
\includegraphics[width=1 \linewidth]{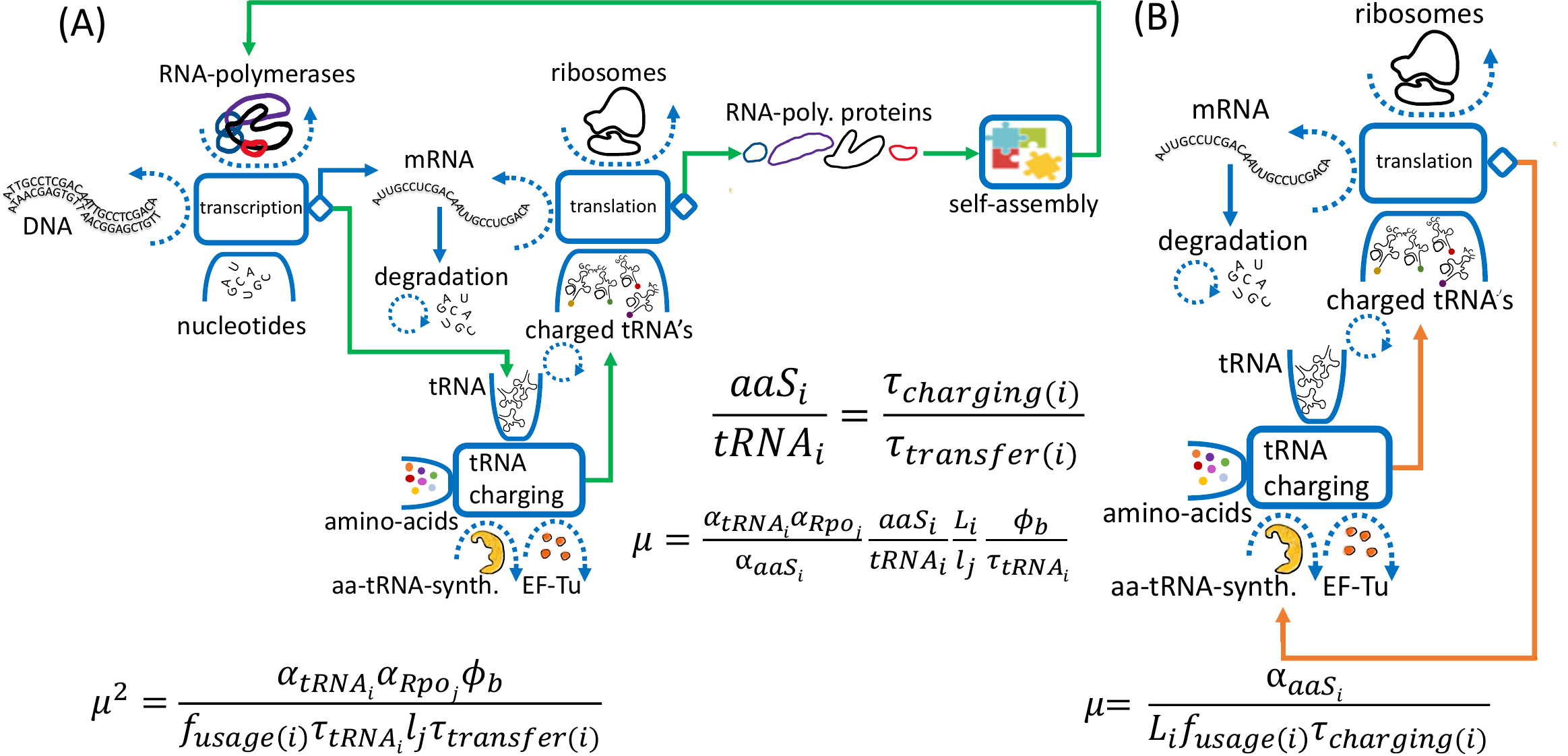}
\caption{tRNA and amino-acyl-tRNA synthetase autocatalytic cycles and growth laws. (A) The tRNA autocatalytic cycle. tRNAs are transcribed by RNA polymerases. After the maturation process (not shown), each tRNA is charged with an amino acid and loaded onto EF-Tu. The EF-Tu-tRNA-aa is also charged with GTP (not shown) and subsequently deliver the amino acid to the ribosome. This cycles repeats until the tRNA degrades in a time scale that we assume to be much longer than the doubling time. Some of the delivered amino acids are embedded in Rpo proteins which self assemble to form new RNA polymerases, some of which are allocated to transcribe tRNAs, thus completing the cycle. (B) The aminoacyl-tRNA synthetase autocatalytic cycle. An aminoacyl-tRNA synthetase protein of a specified type charges tRNA with its corresponding amino acid. A fraction of the charged tRNAs contributes the amino acids to form new aminoacyl-tRNA synthetases of the same type, thus completing the cycle. In the middle of the figure, we show the ratio between tRNA of type $i$ and its associated aminoacyl-tRNA synthetase. Neglecting cross charging, this ratio equals, at balanced growth, to the ratio of the charging cycle duration to the charging duration.}\label{trnafig}
\end{figure}

For the three environments tested in the experiment, we fitted the RNA polymerase allocation parameter to yield the nominal growth rate without limiting the expression of the rpoB and rpoC genes. All other parameters were taken from known measurements \cite{Bremer,bionumbers}; see SI for more details. Next, we fitted the values of $K$ and $h$ to the growth rate as a function of the fluorescence in the M9+glucose medium. We found that the values of both $K$ and $h$ remained valid for the other two environments, without further fitting, supporting the hypothesis that our model is consistent. 

In the second experiment \cite{Jun17},  sub-lethal dosages of rifampicin---a drug that targets DNA-bound RNA polymerases that are just beginning to transcribe RNA---was administered to \textit{E. coli}. The dependence of the growth rate and the RNA-to-proteins ratio as a function of the concentration of rifampicin were both measured. The growth rate was found to decrease as the concentration of rifampicin increased. The RNA-to-protein ratio however, remained constant, in marked difference to the ribosome growth law, where RNA-to-protein ratio was found to change linearly with the growth rate \cite{hwa1}. 
 
We explain this discrepancy by using our RNA polymerase growth law (Eq. \ref{RNAPgl}). Our model naturally accounts for these observations by assuming that rifampicin turns the RNA polymerase autocatalytic cycle to the limiting cycle by reducing the number of active RNA polymerases. As the concentration of rifampicin increases, the fraction of active RNA polymerases decreases and, accordingly, the growth rate of RNA polymerases decreases. Moreover, the decrease in the number of active RNA polymerases globally decreases RNA transcription in the cell, thus reducing the mRNA levels of all proteins, ribosomal and non-ribosomal alike. This process explains why the RNA-to-protein ratio remains constant. We also derived this result by rewriting the ribosome growth law under the assumption that mRNA is limiting (see SI). We note that the RNA-to-protein ratio could potentially increase if protein synthesis is affected more severely than ribogenesis. 

In a growing cell, most RNA transcription is of rRNA. Therefore, it is natural to ask whether rifampicin reduces growth by preventing rRNA transcription and thereby, ribogenesis. Using our model, we can calculate the growth rate and RNA-to-protein ratio for various limitation regimes. We consider three relevant limitation regimes; (i) rRNA is limiting but mRNA is not; (ii) both rRNA and mRNA are limiting; (iii) rRNA is not limiting but mRNA is. 

In the first limitation regime, rRNA can limit the synthesis of new ribosomes because RNA polymerases are limited, but mRNA does not limit translation. The reduction in rRNA synthesis is accompanied by a reduction in ribosomal protein translation. This is due to a remarkable mechanism discovered by Nomura \cite{Nomura1}, namely, that ribosomal proteins that are primary rRNA binders can down-regulate their own translation, as well as the translation of other ribosomal proteins on the same operon, if they fail to find their target rRNA sequence. This translational feedback mechanism is due to an affinity of ribosomal proteins which are primary binders \cite{Nierhaus}, to bind to a region on their mRNA, which is similar to their rRNA binding site. Binding of these proteins to their own mRNA prevents further translation from these mRNAs. Together with other mechanisms, this translational feedback keeps the levels of ribosomal protein precursor pools in sync with rRNA transcription, which, in turn, is governed by the modulation of RNA polymerase transcription via the stringent response \cite{Stringent}. Thus, a limitation on rRNA transcription without an accompanied limitation on mRNA is expected to reduce ribosomal protein translation and, thereby, increase the number of ribosomes that are allocated to translate other proteins. Therefore, the regime of rRNA limitation without mRNA limitation is expected to reduce the RNA-to-protein ratio, in contrast to the observation that it remained constant. 

The second limitation regime to consider is the regime in which both rRNA and mRNA simultaneously limit ribogenesis and translation. This regime is consistent with a constant RNA-to-protein ratio, as the decrease in translation is common to all protein-sectors. In particular, the ribosomes that are freed from making ribosomal proteins cannot synthesize other proteins instead, because mRNA is in shortage. However, a global shortage in mRNA also means that the RNA polymerase autocatalytic cycle is limiting, as the shortage in mRNA is eventually the result of a shortage in RNA polymerase. We thus conclude that, if both rRNA and mRNA are limiting because RNA polymerase is limiting, growth rate will be determined by the RNA polymerase autocatalytic cycle, and the RNA-to-protein ratio will remain constant.

The third limitation regime to consider is that in which rRNA is not limiting and mRNA is limiting. This regime is also consistent with a constant RNA-to-protein ratio, as the global shortage in mRNA implies a reduction in protein synthesis, including of ribosomal proteins. If only mRNA is limiting because RNA polymerase is limiting, growth rate will be determined by the RNA polymerase autocatalytic cycle, and the RNA-to-protein ratio will remain constant.

A possible alternative explanation to the decrease in growth rate as a function of rifampicin concentration is that the mRNA shortage makes a particular metabolic autocatalytic cycle limiting. However, in this case, we would expect the stringent response to reduce the production of rRNA, which, in turn, would reduce the translation of ribosomal proteins. The freed ribosomes would be diverted to translation of proteins belonging to the limiting metabolic cycle. Thus, if a metabolic autocatalytic cycle were limiting, we would expect the RNA-to-protein ratio to decrease with increasing rifampicin concentrations.

\textbf{The tRNA-synthetase and tRNA autocatalytic cycles.} 
To translate mRNAs, ribosomes rely on a series of essential proteins that facilitate ribosome binding to mRNA, tRNA charging and transfer into the ribosome, translocating the ribosome along the mRNA, and releasing the ribosomes from the mRNA upon completing the translation process. 

In all autocatalytic cycles, any of the catalysts that are part of the cycle can be seen as the pivot catalysts, around which the autocatalytic cycle is constructed. To demonstrate this notion, we present two growth laws --- aminoacyl-tRNA synthetase growth law and tRNA growth law (Fig. \ref{trnafig}). 

Consider, first, the aminoacyl-tRNA synthetase ($aaS$) for a particular amino acid $i$. The $aaS_i$ catalyzes the loading of an amino acid of type $i$ onto its corresponding tRNA. The loaded tRNA then binds to EF-Tu---the most abundant protein in \textit{E. coli}---and proceeds to enter a ribosome and deposits the amino acid to the elongating peptide-chain, which, subsequently, folds to form the new protein. A fraction of the proteins formed will be $aaS_i$ proteins, thus closing the cycle. 

Similarly, consider tRNA synthesis by RNA polymerases. After maturation, the transcribed tRNAs are charged with amino acids and  transferred to ribosomes. A fraction of these tRNAs will contribute their amino acids to Rpo proteins, which, in turn, self-assemble to form new RNA polymerases, some of which are allocated to make tRNAs, thereby closing the cycle. More generally, using this method allows us to connect almost any two processes in the cell e.g. transcription, translation, tRNA charging, metabolic rate, DNA synthesis rate, membrane synthesis and assembly times of protein complexes. 

In the current context of tRNA charging, we find a connection between the number of tRNAs and the number of RNA polymerases required to sustain growth at a given rate $\mu$, given the allocation of RNA polymerase toward the transcription of tRNA and rpo genes. We find that the growth rate $\mu$ is equal to the fraction of active RNA polymerases transcribing mRNAs divided by the transcription duration multiplied by the active RNA polymerase to tRNA ratio, $\mu = \frac{\alpha_{tRNA}}{\tau_{tRNA}} \times \frac{Rpol_b}{tRNA}$.

Traditionally, the ratio of tRNA to ribosomes was studied since tRNA transports amino acids to the ribosomes. Additionally, RNA polymerase to ribosomes ratio was studied since RNA polymerase writes mRNA instructions for ribosomes to translate. 
However, evidently, tRNA also serves RNA polymerases, albeit indirectly, because RNA polymerases are made of proteins and tRNA had to deliver the amino acids required for making these proteins to the ribosomes that were synthesizing them. Thus it also makes sense to inquire about the ratio between tRNAs and RNA polymerases, at balanced growth conditions. 

The amino-acyl-tRNA synthetase growth law that we derive is given by 
\begin{align}
\mu = \frac{\alpha_{aaS_i}}{L_i f_{usage(i)} \tau_{charging(i)}},
\label{tRNA1}\end{align}
where as above, $\mu$ is the growth rate, $\alpha_{aaS_i}$ is the fraction of active ribosomes translating $aaS_i$ proteins, $L_i$ is the number of amino acids in $aaS_i$, $f_{usage(i)}$ is the fraction of amino acids of type $i$ used in the entire proteome and $\tau_{charging(i)}^{-1}$ is the rate at which $aaS_i$'s are charging their corresponding tRNAs.

The tRNA growth law that we derive is given by
\begin{align}
\mu^2 = \frac{\alpha_{tRNA_i} \alpha_{Rpo_j} }{f_{usage(i)} \tau_{tRNA_i} l_{j} \tau_{transfer(i)}} \tilde{\phi}_b
\label{tRNA2}\end{align}
where $\alpha_{tRNA_i}$ is the fraction of active RNA polymerases transcribing $tRNA_i$,  $\alpha_{Rpo_j}$ is the fraction of active ribosomes translating the $j^{th}$ $Rpo$ protein, $l_{j}$ is the number of amino acids in $Rpo_j$, $\tau_{transfer(i)}$ is the charging cycle duration of $tRNA_i$, i.e., the average duration between two subsequent charging events, and $\tilde{\phi}_b$ is the fraction of active RNA polymerases.

Since Eq. \ref{tRNA1} and Eq. \ref{tRNA2} are coupled via the tRNA charging reaction, we can reduce the order of Eq. \ref{tRNA2} by  substituting for $f_{usage(i)}$, derived from Eq. \ref{tRNA1}, in Eq. \ref{tRNA2} to obtain
\begin{align}
\mu=\frac{\alpha_{tRNA_i} \alpha_{Rpo_j}}{\alpha_{aaS_i}} \frac{\tau_{charging(i)}}{\tau_{transfer(i)} \tau_{tRNA_i}} \frac{L_i}{l_j}.
\label{yagl}
\end{align}
Where we assume, for simplicity, that all RNA polymerases are active, i.e., that $\tilde{\phi}_b$=1 (otherwise, right hand side of Eq. \ref{yagl} should be multiplied by $\tilde{\phi}_b$).

We also find that the ratio between the tRNA and its aa-tRNA-synt. is inversely proportional to the rate of tRNA amino acid transfer cycle to the rate of aa-tRNA-synt charging,
\begin{align}
\frac{tRNA_i}{aaS_i} =\frac{\tau_{transfer(i)}}{\tau_{charging(i)}}.
\end{align}
Thus, the faster the charging of a particular aa-tRNA-synt., the fewer tRNAs per aa-tNRA-synt. are required. 
Ttesting the prediction arising from these tRNA growth laws, requires accurate measurements of the fraction of active RNA polymerase transcribing each gene, and of the tRNA abundances and the growth rate, all under the same conditions and, preferably, on the same experiment.

In Fig. \ref{trnafig}, we present these two autocatalytic cycles and the resulting growth laws.

\subsection*{Further applications of the autocatalytic growth laws}
\textbf{Growth rate dependence on temperature.}
Our starting point is the bacterial growth law $\mu=\gamma_{transl.}(\Phi_R-\Phi_0)$ which was experimentally measured at $37 ^oC$. We recall a few experimental facts. The first observation is the existence of an Arrhenius regime, which is a range of temperatures between $T_{Ac}=20 ^oC$ and $T_{Ah}=40 ^oC$ in \textit{E. coli}, where the RNA-to-protein ratio does not change while the ribosome elongation rate changes with an Arrhenius temperature dependence, i.e., $\hat{\gamma}_{elong.}= \gamma_{elong.}e^{\frac{\Delta G}{k_B T}}$. The fact that growth rate scales with temperature with an Arrhenius type dependence, which is typically relevant to a single chemical reaction, might seem surprising; however, if the ribosome autocatalytic cycle is the limiting cycle, all other autocatalytic cycles locks to its growth rate and the scaling becomes a natural consequence of the increase in the elongation rate with temperature. Furthermore, the fact that the ribosome fraction remains constant within the Arrhenius regime comes as a natural consequence of the fact that this cycle is the leading cycle across the entire Arrhenius regime. What happens beyond the Arrhenius regime? both above and below this regime, the cell cannot sustain growth, because an increasing number of proteins denature, if $T>T_{Ah}$, or misfold, if $T<T_{Ac}$, and more ribosomes need to be allocated to mitigate the loss of functioning proteins. 

It is possible, although not essential to assume, that metabolic proteins are the first to cause such a reallocation of ribosomes, due to the stringent response. We account for this reallocation by assuming a linear change in the ribosome fraction, from $\phi_R$ at the hot $T=T_{Ah}$ and cold $T=T_{Ac}$ edges of the Arrhenius regime, down toward $\Phi_0=3.5\%$ which is the minimal allocation at zero growth rate. We use data that show that \textit{E. coli} ceases to grow at $T_{cd}=8 ^o C$ and at $T_{hd}=49 ^o C$ \cite{bionumbers}. 

To account for the growth rate dependence on temperature, we make minimal modifications to the ribosomal growth law \cite{hwa1}, making it consistent with the above mentioned observations. First, we scale the elongation rate for all temperatures by the Arrhenius factor, $\hat{\gamma}_{elong.}= \gamma_{elong.}e^{\frac{\Delta G}{k_B T}}$. For the Arrhenius regime, we write $\mu=(\Phi_R-\Phi_0) \gamma_{elong.}e^{\frac{\Delta G}{k_B T}}$, where $\Phi_R$ is the ribosome fraction at the maximal growth conditions at $37 ^o C$, $\Phi_R \approx 0.3$ \cite{hwa1}. Finally, for temperatures from $T_{Ah}$ up to $T_{hd}$, and from $T_{Ac}$ down to $T_{cd}$, we scale the ribosome fraction $\Phi_R$ linearly from its maximal value down to $\phi_0 \approx 0.035$, which is the allocation below which growth is brought to a halt. 
\begin{figure}[t]
\centering
\includegraphics[width=1 \linewidth]{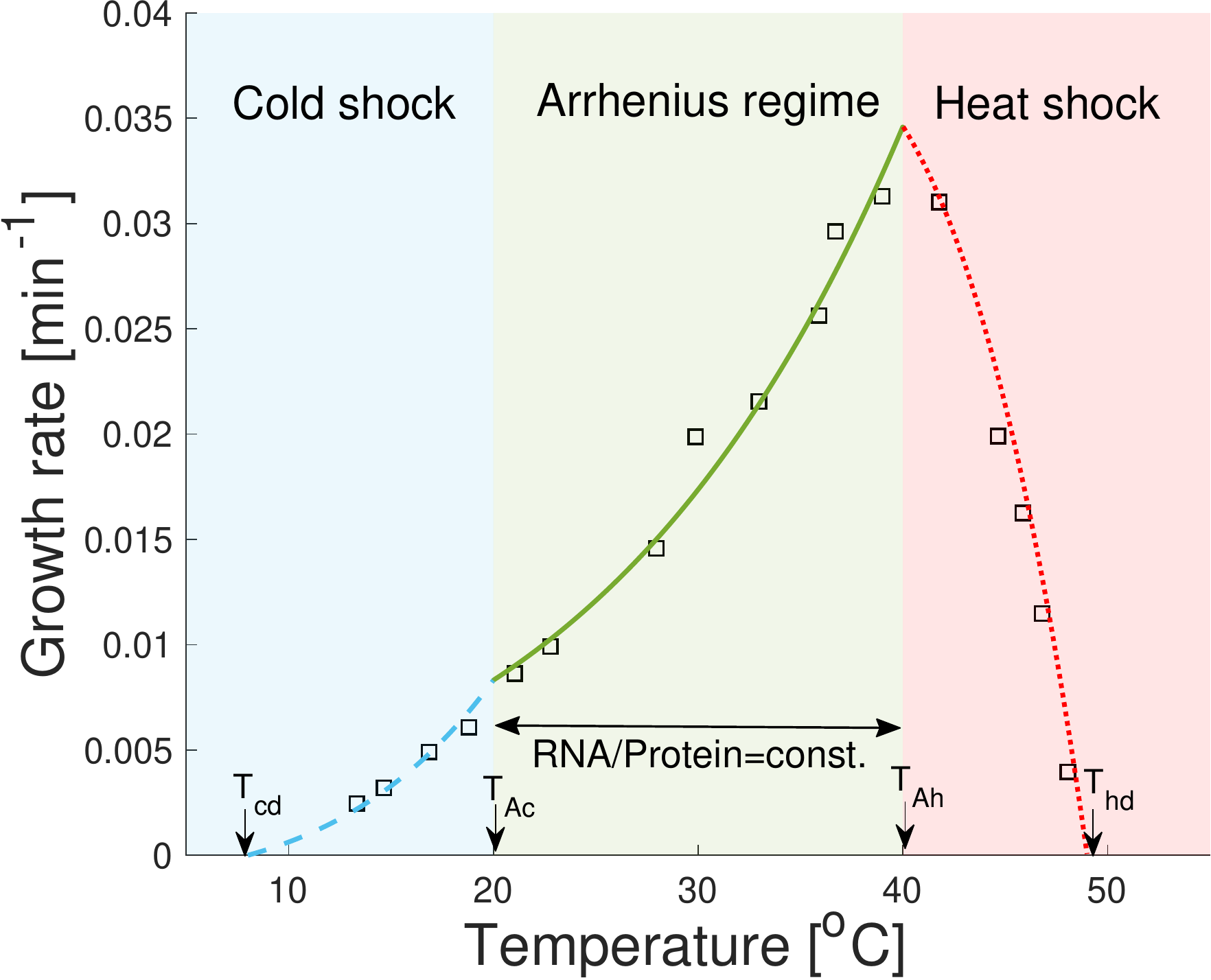}
\caption{Growth rate dependence on temperature.  Four temperatures, derived from experiments, are required to account for the observations: the cold temperature $T_{cd}=8  ^o C$, at which growth rate ceases; the temperature range $[T_{Ac},T_{Ah}]=[20,40]  ^oC$, which defines the Arrhenius regime, where the RNA-to-protein ratio is measured to be constant \cite{Arrenhius};  the hot temperature $T_{hd}=49  ^oC$, at which growth ceases, i.e., $\mu=0$;  and the ribosome elongation rate dependence on the temperature, which is assumed to change by the Arrhenius factor $\gamma_{elong.}(T)=\gamma_{opt}e^{\frac{\Delta \tilde{G}}{k_B T}}$, where $\Delta \tilde{G}$ is calculated based on a model presented in \cite{dill3} as explained in the main text. Beyond the Arrhenius regime, we assume the RNA-to-protein ratio decreases linearly from its nominal value at both sides of the Arrhenius regime to its minimal value $\Phi_0=0.035$, both at $T_{cd}$ and at $T_{hd}$.}
\label{muvst}
\end{figure}
\[
    \mu(T)= \gamma_{elong.} e^{\frac{\Delta \tilde{G}}{k_B T}} \times
\begin{cases}
       \Delta \Phi \frac{ (T-T_{Ac})}{(T_{Ac}-T_{cd})}+\Delta \Phi, T<T_{Ac}\\
\Delta \Phi \hspace{2.2 cm},T\in [T_{Ac},T_{Ah}]\\
    \Delta \Phi \frac{ (T-T_{Ah})}{(T_{Ah}-T_{hd})}+\Delta \Phi, T>T_{Ah} \\
\end{cases}
\]
where $\Delta \Phi =\Phi_R-\Phi_0$, $\Phi_R$ is the measured ribosome fraction in the medium at $37 ^oC$ and $\Phi_0$ is the ribosome fraction when growth rate is zero, taken to be a constant independent of the media, as measured in \cite{hwa3}. The elongation rate in the medium at $T=37 ^o C$ is given by $\gamma_{elong.}$ and $\Delta \tilde{G}= \Delta G(T) - \Delta G(T=37 ^o C)$. 

Following \cite{TempFBA}, we calculate $\Delta G(T)$ by using the formula developed in \cite{dill3}, namely,
\begin{align}
\Delta G(T)=\Delta H+\Delta C_p \left(T-\hat{T} \right)-T \Delta S-T \Delta C_p ln(\frac{T}{385}),
\end{align}
where $\Delta H=4N+143$, $1000 \times \Delta S=13.27N+448$, $\Delta C_p=0.048N+0.85$, and $\hat{T}=49 ^oC$. We calculate the average expressed protein size $N=364$ amino acids from the ribosome profiling experiment presented in \cite{RibosomeProfiling2}, for MOPS rich media at $37 ^oC$ (doubling time of $21$ min).
In Fig. \ref{muvst}, we plot the predicted growth rate as a function of the temperature. We obtain a good fit to the measured temperature dependence of the growth rate \cite{bionumbers}, without using fitting parameters. 

\textbf{Effect of lamotrigine on ribosome assembly duration in \textit{E. coli}.}
As presented in Eq. 1, the ribosome growth law can be used to yield the relationship between growth rate and all the time-scales in the cycle, namely, the ribosome resting duration $\tau_0$, the translation duration of all ribosomal proteins $\tau_R$, the assembly duration $\tau_{SA}$, and the allocation parameter $\alpha_R$, which represents the percentage of ribosomes that actively translate ribosomal proteins. Upon the approximation that $\mu \tau_0 \ll 1$, we obtain a second order equation for the growth rate, which is readily solved,
\begin{align}
\mu = \frac{\sqrt{1+\alpha_{RP_i} \frac{4\tau_{SA}}{\tau_{RP_i}}} - 1}{2\tau_{SA}},
\label{ribosomeSALaw}
\end{align}
where, as above, $\tau_{SA}$ is the ribosome assembly duration, $\tau_{RP_i}$ is the translation duration of a ribosomal protein that is a primary rRNA binder, and $\alpha_{RP_i}$ is the fraction of active ribosomes allocated to translating this ribosomal protein. 

By using a Taylor expansion of the numerator in Eq. \ref{ribosomeSALaw} we obtain
\begin{align}
\mu \approx \frac{\alpha_{RP_i}}{\tau_{RP_i}} - \frac{\alpha_{RP_i}^2 \tau_{SA}}{\tau_{RP_i}^2} + \ldots,
\label{ribosomeSALawTaylor}
\end{align}
which to first order is $\mu \approx {\alpha_{RP_i}}{\tau_{RP_i}^{-1}}$, i.e., to first order the ribosome assembly time does not affect the growth rate. This first order term is identical to the bacterial growth law when setting $\Phi_0=0$. For $\frac{\alpha_{RP_i} \tau_{SA}}{\tau_{RP_i}} \ll 1$, the second order correction to the growth rate is negligible. However, when $\frac{\alpha_{RP_i} \tau_{SA}}{\tau_{RP_i}} \sim 1$, the growth rate will decrease towards zero as the assembly duration increases.

\begin{figure}[ht]
\centering
\includegraphics[width=1 \linewidth]{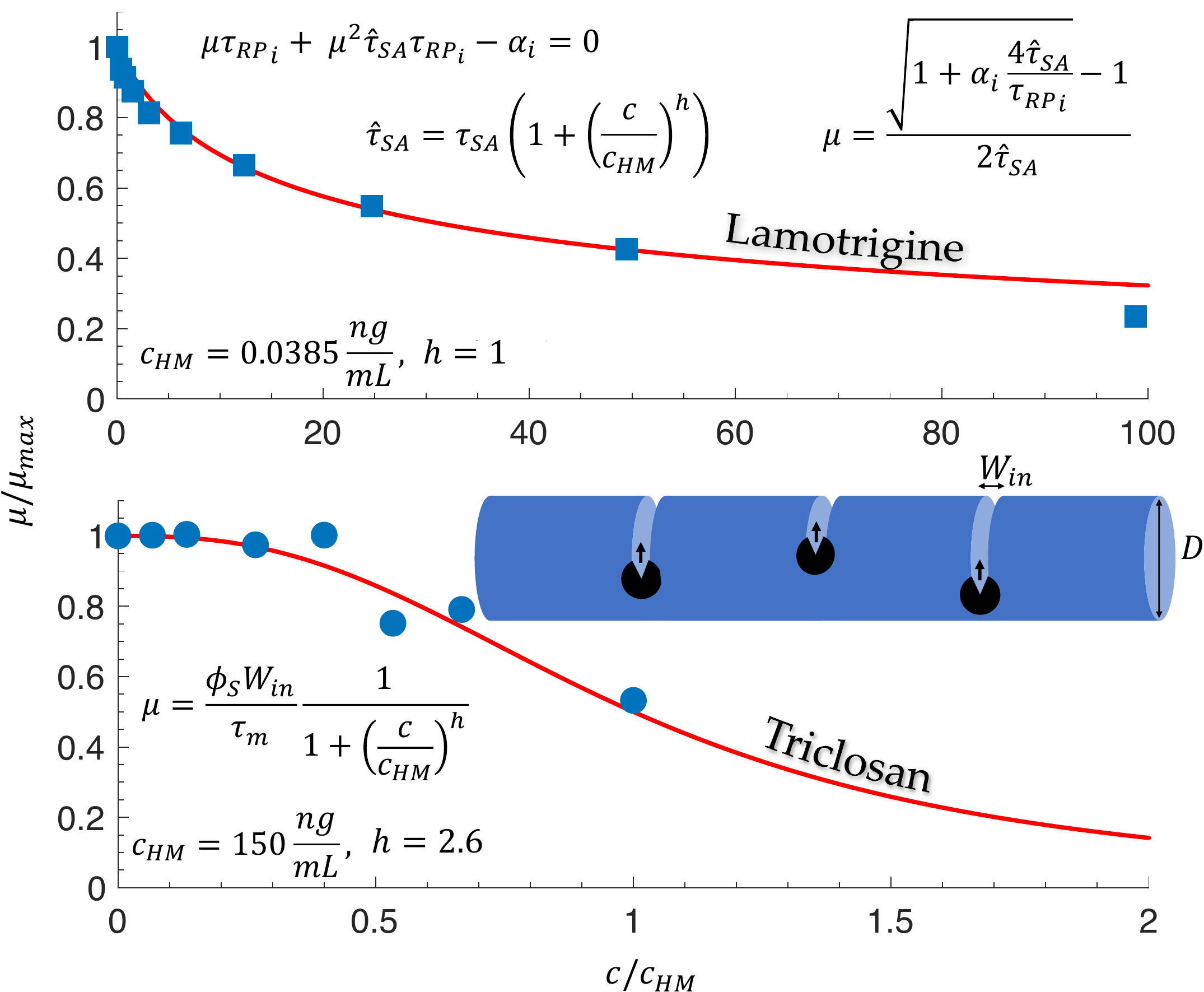}
\caption{Effect of two antibacterial agents---lamotrigine and triclosan---on \textit{E. coli} growth rate.  Upper panel depicts the effect of lamotrigine, an anti-convulsive drug that was recently found to adversely affect ribosome assembly in \textit{E. coli} \cite{Lamotrigine}. The growth law that we use to fit the data is based on Eq. 1 in the main text. We neglect both the ribosomal protein pool time and the free ribosome idling time to obtain a simple quadratic equation. When the assembly duration becomes comparable or longer than the bare doubling time $\sim \frac{\tau_{RP_i}}{\alpha_i}$, non-negligible corrections to the standard bacterial growth law are required. In the lamotrigine experiment this occurs at concentrations above $0.03$ $ng/mL$. The lower panel depicts the effect of triclosan on the growth rate of E. coli. Data were taken from \cite{Jun17}. Triclosan disrupts the protein FabI, which catalyzes an essential step in the biosynthesis of fatty acids required for building the bacterial membrane. To incorporate this effect, we decrease the utilization of the membrane synthesis proteins (depicted in black in the illustration above the graph). This is a simplification because there are several such proteins that collectively synthesize the membrane. The membrane synthesis rate $\tau_{m}^{-1}$ is unaffected by triclosan at concentrations  $c < c_{HM}=0.15$ $\mu g/mL$.  Administering triclosan reduces the size of the pool, until, at a critical concentration, the pool is depleted. Thereafter, any delay in the supply of fatty-acid precursors delays the insertion of material into the elongating membrane by the membrane synthesis proteins. Thus, after the critical concentration $c_{HM}$ is surpassed, the growth rate starts to decrease. To accommodate this switch-like behavior, we used a Hill function with two fitting parameters, $K=0.15$ and Hill's coefficient $h=2.6$. Schematic figure depicts three membrane synthesis protein complexes $C=3$, inserting new membrane patches of width $W_{in}$ into the existing membrane. The growth law equates the rate of growth $\mu$, the width of the inserted patch $W_{in}$, the surface concentration of the membrane synthesis proteins $\phi_S$, and the insertion rate $\tau_{m}^{-1}$.} 
\label{fig:5}
\end{figure}

Recently, the anti-convulsion drug lamotrigine was found to adversely affect the ribosome assembly process in \textit{E. coli} \cite{Lamotrigine}. Using the general form of the ribosome growth law that accounts for assembly time we can fit the observed dependence of the growth rate on the concentration of lamotrigine. This is done by equating $\hat{\tau}_{SA} = \tau_{SA} \frac{1}{1+(\frac{c}{c_{HM}})^h}$, and  then fitting $c_{HM}$ and $h$. We find that $c_{HM}=0.0385$ ng/mL which is in accord with the empirical concentration at which the growth rate is reduced by half. We also find that $h=1$, indicating a non-cooperative effect of lamotrigine on the ribosome self-assembly process. In Fig. \ref{fig:5}, we present the resulting fit. A more detailed theory is required in order to obtain a mechanistic understanding of the effect of lamotrigine on the assembly process, which we leave for a future study.

\textbf{Effect of triclosan on cell-wall synthesis in \textit{E. coli}.} Cylindrically shaped bacteria, such as \textit{E. coli}, are known to elongate at an exponential rate \cite{Jun10}. If the width of the bacteria does not change significantly over the doubling time, this also implies exponential growth of the volume and surface area. However, the process of synthesizing new membrane is linear: the membrane is inserted through an insertion site in a process hypothesized to be coordinated by proteins belonging to the ``elongasome'', including MreB and penicillin-binding proteins that serve key roles, which are still subject to research \cite{memb1,memb2}. 

A mechanism that can coarsely account for the observation of exponential elongation is asynchronous threshold initiation of new insertion sites by constitutive membrane synthesis proteins. Under balanced growth conditions, these proteins are expected to grow exponentially due to the inherent coupling with the transcription--translation machinery that synthesizes them. If either the initiation is asynchronous or there is an inhomogeneity in the rate of insertion between different insertion sites, a smooth exponential growth of the membrane ensues.  

We devised a simple mathematical description of this coupling, and obtained the following relative abundance growth law, which relates the growth rate $\mu$ to the surface density of insertion sites $\phi_{S}$, the width of the insertion $W_{in}$, and the speed of insertion in units of length over time $\tau_m^{-1}$:
\begin{align}
\mu = \frac{\phi_{S} W_{in}}{ \tau_m}. \label{eqMem}
\end{align}
This equation is identical to the equation used in \cite{memb2} to obtain the width of insertion site as a function of the growth rate.
 
Membrane bound volume and membrane surface area are also physical resources that are essential to all cellular processes simply because all cellular constituents occupy some volume. It is also required in order to maintain the reaction rates and prevent dilution by diffusion. Furthermore, surface area is required by metabolism in order to exchange metabolites and heat with the surrounding environment \cite{realestate,hwa2}. Obtaining a close cycle growth law is, however, more involved than the other cycles we already discussed and goes beyond the scope of this work.

What happens when membrane synthesis is disrupted? consider the antibacterial agent triclosan, which targets fatty-acid biosynthesis. Triclosan disrupts the protein FabI, which catalyzes an essential step in the biosynthesis of fatty-acids, necessary for building the bacterial membrane. To explain the effect of triclosan on the growth rate we modulate the rate of insertion of new membrane surface area in Eq. \ref{eqMem} by a Hill function, such that the insertion rate is monotonically decreasing with the triclosan concentration. This accounts for the decrease in the speed of insertion caused by the dwindling of the fatty acid supply. Thus, the new membrane synthesis rate is $\hat{\tau}_m^{-1} =\tau_m^{-1} \frac{1}{1+\frac{c}{c_{HM}}}$. Using data from \cite{Jun17}
 we found by fitting that $c_{HM}=150$ ng/mL and $h=2.6$. See inset A in figure \ref{fig:5}. 

\section*{Discussion}
The existence of simple mathematical relationships between growth rate to the rate of cellular processes, such as, transcription, translation, metabolism, and membrane synthesis seems to suggest that understanding the causal relations between growth rate and these processes is also at hand. Unfortunately, this is not the case. 

The two types of growth laws obtained per autocatalytic cycle---the closed-cycle growth law and the relative abundance growth law---require local knowledge about the time scales and relative abundances of catalysts within the cycle. In addition, they require global knowledge regarding the allocation parameters of the various catalysts, e.g., the fractions of RNA polymerase and ribosomes that are dedicated to transcribing and translating the relevant proteins. Understanding the causal relation between these variables and the growth rate requires a global understanding of how the allocation parameters are determined. This understanding, in turn, requires knowing how the cell allocates its resources among all its autocatalytic cycles and, in particular, understanding the evolutionary design-logic, behind the transcription, translation and metabolic control mechanisms.

The ribosome bacterial growth law also do not provide a causal explanation. This is because understanding how the ribosome fraction and the elongation rate change when the nutritional composition of the environment changes requires a causal model that explains how the cell controls the global allocation of its resources among all coupled autocatalytic cycles. In the absence of such a mechanistic model, we cannot distinguish between different mechanisms that lead to the same growth rate. For example, in harsh environments, the cell could respond by modifying the allocation towards ribosomal proteins, while keeping the rate of translation fixed. Alternatively,  the cellular response can be to modify the allocation toward ribosomal proteins in a manner that will reduce the elongation rate, as compared with a permissive environment. Distinguishing between these two possibilities requires a measurement of the elongation rate and ribosome fraction together \cite{hwa3}, while a causal model would be able to predict the interdependence between the elongation rate and the ribosome fraction, as a function of the growth rate.

While much is already known about transcription, translation and metabolic control mechanisms, a holistic picture is yet to form. A few general principles do seem to emerge, such as allosteric regulation to control gene expression in response to changes in external metabolite composition \cite{Sauer} and product feedback inhibition, e.g., in metabolism, which enable the cell to optimally tune gene-expression and protein content \cite{FeedbackInhibition}.  

The success of bacterial growth laws in describing the relationship between the growth rate and different physiological parameters lays on two pillars. The first pillar is the fact that typically autocatalytic network yields exponential balanced growth. The second pillar is the universal structure of the transcription--translation autocatalytic network, and the universal manner in which it couples to other cellular autocatalytic cycles, leading to balanced exponential growth and to many biological constraints between different physiological parameters.

In the clash between the physics inspired strive for simple underlying laws of bacterial physiology, to the biological hard-won understanding of the intricacies of life, we end in a middle-ground. On one hand, we have found valid and simple growth laws. On the other hand, we demonstrated that the validity of a given growth law do not fully reveal the physiological state of the cell. Understanding how the cellular state is determined in response to internal and external cues, and how evolutionary stresses shaped different schemes for determining it, remains a formidable challange.

\section*{Methods}
A standard approach to modeling kinetics is to focus on concentrations and write ordinary differential equations (ODEs) for their rate of change.  A well-known example is the Michaelis-Menten form $\frac{dc}{dt}=\frac{v_m c}{K+c}$, which is non-linear. 
\begin{figure}[h]
\centering
\includegraphics[width=1 \linewidth]{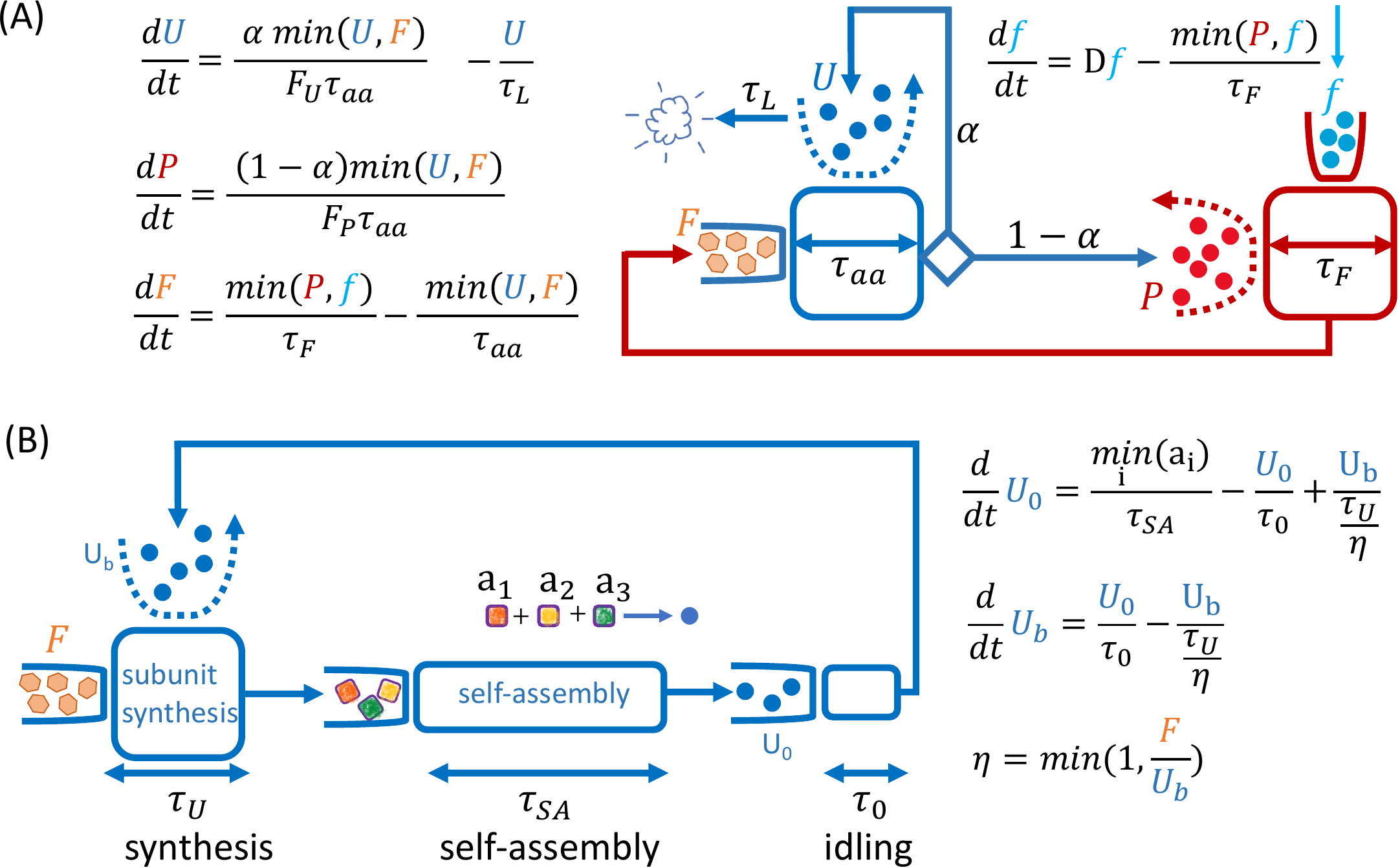}
\caption{Two-step model buildup. (A) Coupling of two autocatalytic cycles (compare with \cite{dill2}). Two minimum functions imply four different limitation regimes. (B) Adding idle and active states and an intermediate self-assembly step. The equations provide a quantitative description of the graphical notation.}
\label{FigMethods}
\end{figure}

\begin{figure}[h]
\centering
\includegraphics[width=1 \linewidth]{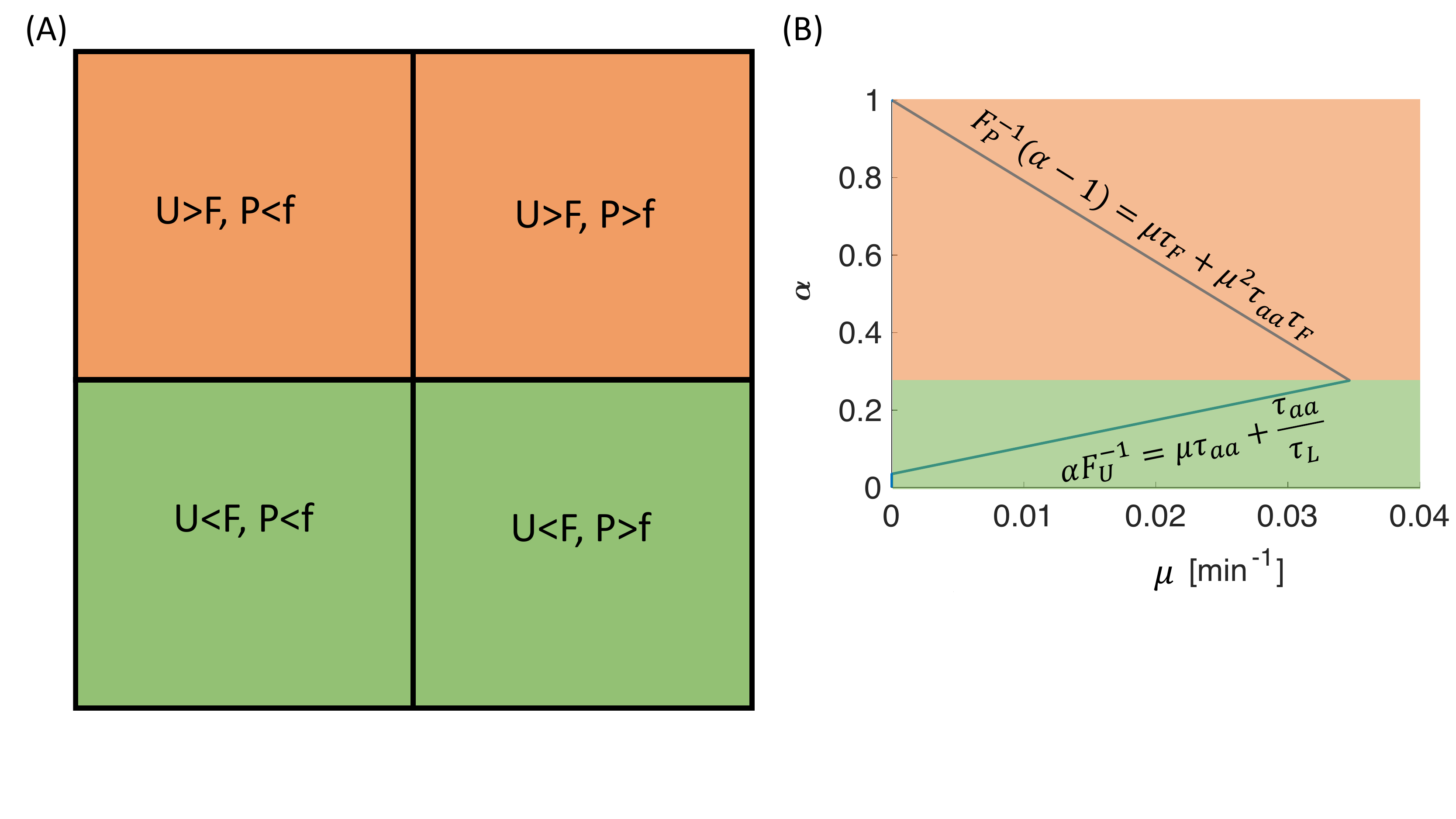}
\caption{Limitation regimes and growth regimes for the ``UPF model'' presented in upper panel of figure \ref{FigMethods}. Panel A presents the four different limitation regimes. Different limitation regimes which share a limiting cycle belong to the same growth regime. The growth regime is colored green for $U < F$, i.e., when the $U$ machines are not starved for $F$ substrates, and are colored orange when $U > F$, i.e., when the $U$ machines are starved for $F$ substrate. Panel B presents the allocation parameter as a function of the growth rate, using the two growth laws derived in the Methods section in the main text for the two growth regimes.}
\label{last}
\end{figure}

Instead of using coupled non-linear ODEs for concentrations, we focus on the rate of increase of the absolute number of molecules. This approach has two main advantages over other approaches, which are also approximations. First, by counting molecules, exponential growth---the hallmark of autocatalysis---becomes evident.  Second, catalysts in the broader sense of the word, i.e., materials that are required to facilitate a reaction but are not consumed by it, and substrates, i.e., materials required by a reaction and are consumed by it, acquire equal footing in the equations. This facilitates a straightforward characterization of which cycle is limiting and which is not, under different circumstances. 

We begin by explaining how we translate the graphical notation presented in Figs. \ref{fig1} and \ref{fig2} into a set of coupled ODEs, which are piecewise linear, and how we derive from them various growth laws that are valid for different limitation regimes---which we also define. 

For the sake of clarity, we use a simplified toy model to elucidate the two-step process that we employ to devise the more complex models, which bear biological relevance and are presented in the main text. The first step was to identify all autocatalytic cycles and the second step was to insert more internal states per autocatalytic cycle to account for idling vs. active catalysts, self-assembly steps, and precursor pools, as we explain further below. 

We recall the proverb attributed to G. Box, ``All models are wrong, but some are useful''. We argue that the usefulness of our modeling approach lies in its modularity, and in its ability to unravel both simple growth laws and the complex circumstances that renders them valid.
  
Consider then the drastically simplified model shown in Fig. \ref{FigMethods}A, which we term the ``UPF model''. In this model, a fraction $\alpha$ of machines of type $U$ catalyze themselves. The remaining $U$ machines synthesize a second type of machines $P$. The $P$ machines convert an external substrate $f$ to an internal substrate $F$, which is used by $U$ to make both more copies of itself and new $P$s. The $U$ machines have a lifetime $\tau_L$, while the $P$ machines have an infinite lifetime. To catalyze a new $U$ machine, $F_U$ units of $F$ are required. To catalyze a new $P$ machine, $F_P$ units of $F$ are required. 

The rate at which $U$ catalyzes either $P$ or $U$ is equal to the incorporation rate of a single $F$, $\tau_{aa}^{-1}$, times $F_P$ or $F_U$ respectively, assuming that $F$ is abundant. The rate at which $P$ converts an external substrate $f$ to an internal substrate $F$ is $\tau_F^{-1}$, assuming that $f$ is abundant. The minimum function, $min(\cdot)$, in the equations presented in Fig. \ref{FigMethods}A, allows us to uncover the four different limitation regimes of this model. In regime (1), $f \geq P$ and $F \geq U$, both $P$s and $U$s catalyze at their fastest rate, because at any given moment neither of them are starved for substrates. In regime (2) $f \geq P$ and $F<U$, $P$s are not starved for their substrates but $U$s are starved for their substrates. Consequently, the rate at which $U$s operate is reduced by a factor $\eta=\frac{F}{U}$. In regime (3), $f< P $ and $F>U$, $P$s are starved but $U$s are not. Finally, in regime (4), $f<P$ and $F<U$, both $P$s and $U$s are starved for their respective substrates. 

It is important to define the rate of change $\frac{df}{dt}$ and the initial condition $f(t=0)$ of the external substrate $f$ to properly model the experimental circumstances. For example, growth in a chemostat would be modeled by an exponentially growing $f$ over time, with a fixed growth rate $D$ that is equal to the dilution rate of the chemostat, where $D$ is strictly less than the maximally attainable growth rate $\mu_{max}$. In this regime and at steady growth conditions, $f<P$. In contrast, to model turbidostat conditions, $f(t)$ will grow exponentially at a given rate $D \geq \mu_{max}$, and $f \geq P$. To model batch culture growth, $D$ is set to zero, while $f(t=0)=f_0>0$.

To find the growth rate as a function of the time-scales in the model, it is instructive to formulate the equations by using matrix algebra. Define the column state vector $\vec{S}=(U,P,F,f)^{\dagger}$. Then, the dynamics per limitation regime indexed by $i=1, \ldots, 4$ can be described as $\frac{d \vec{S}}{dt} = M_i \vec{S}$. For example, the matrix for regime (1), $M_1$, is defined as $(M_1)_{11}=\frac{\alpha}{F_U \tau_{aa}}-\frac{1}{\tau_L}$, $(M_1)_{21}=\frac{1-\alpha}{F_P \tau_{aa}}$, $(M_1)_{31}=-\frac{1}{\tau_{aa}}$, $(M_1)_{32}=\frac{1}{\tau_F}$, $(M_1)_{42}=-\frac{1}{\tau_F}$ and $(M_1)_{44}=D$, while all other elements of $M_1$ are zero. We use absorbing boundary conditions for all the state variables, to ensure their non-negativity: $\vec{S} \geq 0$. 

Since for each limitation regime the matrix $M_i$ is constant, the coupled ODEs become linear and thus support exponential growth. At steady growth conditions, the solution will be $\vec{S}(t) = \vec{S}_i e^{\mu t}$, where $\mu$ is the largest eigenvalue of $M_i$, and $\vec{S}_i$ is the corresponding eigenvector, i.e., $M_i \vec{S}_i = \mu \vec{S}_i$. In this model, there are two coupled autocatalytic cycles, ``Us make Us'' and ``Ps make Fs, Fs make Ps''. Evidently, all components of both cycles will grow at the same growth rate, irrespective of the value of the allocation parameter $\alpha \in (0,1)$ that couples the two. The same mathematical structure repeats for more complex models (see SI).

Both regime (1) and regime (3), for example, share the same growth law, which can be analytically calculated to be $\mu=\frac{\alpha}{\tau_U} -\frac{1}{\tau_L}$. We define the union of these two limitation regimes as a ``growth regime''. More generally, a growth regime is defined as the union of all limitation regime that share a common limiting autocatalytic cycle. In our case, the common limiting autocatalytic cycle is the ``Us make Us'' autocatalytic cycle. Knowledge that a certain growth law describes a given experimental condition, therefore, is not enough for specifying which limitation regime the cell was in.
If we can measure the parameters of our model, namely the reaction rates and allocation parameters, we can infer the limitation regime by solving for the eigenvectors corresponding to the largest eigenvalue, for all the limitation regime matrices $M_i$ and checking for their consistency. Consistency checking a such eigenvector of a given matrix $M_i$ requires checking if the eigenvector's elements obey the same inequalities as the limitation regime that defined the matrix $M_i$. If all the inequalities are satisfied, the limitation regime $i$ is consistent with the measured parameters. In this case all the other limitation regimes will be inconsistent. For example, in the current simplified model, setting $F_U=F_P=1$, $\tau_U=\tau_F=6$ min, if $\alpha=0.2$ we can show that regime (1) and (3) are consistent, while (2) and (4) are inconsistent. If we further know that, e.g., $D>\mu_{max}$, then only regime (1) will be consistent. 

To yield a positive growth rate, $U$s must autocatalyze at a rate that is faster than their decay rate $\frac{1}{\tau_L}$. This implies a minimal value for the allocation parameter $\alpha$, above which exponential growth begins. In the regimes where $U$s are starved, the second autocatalytic cycle---the ``Ps make Fs, Fs makes Ps'' cycle---determines the growth rate. In this regime, the largest eigenvalue can be again calculated analytically and is given as the positive root of the quadratic characteristic equation, $\mu=\frac{1}{2\tau_{aa}} (\sqrt{1+\frac{4(1-\alpha)\tau_{aa}}{F_P \tau_F}}-1)$. 

Assuming $f>P$ the simple growth law $\mu=\frac{\alpha}{F_U \tau_{aa}} - \frac{1}{\tau_L}$ will be valid as long as $\alpha \leq \alpha_{opt}$ --- the optimal allocation that leads to maximal growth. $\alpha_{opt}$ can be found by equating the two growth laws $\mu=\frac{\alpha}{F_U \tau_{aa}}$ and $\mu=\frac{1}{2\tau_{aa}}(\sqrt{1+\frac{4(1-\alpha)\tau_{aa}}{F_P \tau_F}}-1)$, and solving for $\alpha$. When $\alpha > \alpha_{opt}$ limitation regime (1) eventually becomes inconsistent. 

In actual experiments, where a parameter equivalent to the parameter $\tau_{aa}$ is measured, there is no a-priori way of knowing in which growth regime the measurement was performed. Hence, what is actually being measured may not be the ``bare'' $\tau_{aa}$, i.e., the fastest elongation duration, but rather $\tau_{aa}'$ which can be different due to a hidden utilization factor $\eta \in [0,1]$ that measures the average percentage of time a catalyst is waiting for substrate, at steady growth conditions, i.e., $\tau_{aa}'=\frac{\tau_{aa}}{\eta}$. Using this notion of utilization parameters, $\eta$, we can extend the validity of growth laws beyond the limitation regime that defines them. For example, in the growth regime where $F<U$, we can still write the growth law in the same mathematical form as the growth law for the regime $F>U$: $\mu=\frac{\alpha}{\tau_U'} - \frac{1}{\tau_L}$, with $\tau_U'=\tau_U/\eta$, and $\eta=\frac{\tilde{F}}{\tilde{U}}$, where $\tilde{F}$ and $\tilde{U}$ are the eigenvector components of the matrix in the growth regime $F<U$, colored in orange in Fig. \ref{last}. As mentioned, $\eta=\frac{\tilde{F}}{\tilde{U}}$, is the fraction of time the $U$ machines are waiting for substrate in the regime $F<U$. Of course in the regime $F>U$, $U$ do not wait and $\eta=1$. Extending this idea to the ribosome growth law, (Eq. \ref{eqRpACtot} and Eq. \ref{eq2}), we can expect that whenever ribosomes are starved for charged tRNA, the ribosome growth law will reman valid, as long as the elongation rate of the ribosomes used in the formula is the measured elongation rate rather than the maximal elongation rate. As long as the ribosomes are translating at the maximal rate, they are in one growth regime, equivalent to the $F>U$ regime. When ribosome are starved for charged tRNAs, the elongation rate decreases, and the ribosome growth law remains valid by using the measured elongation rate which is less than the maximal ($\eta<1$). 

The growth law obtained from solving the characteristic polynomial is the closed-cycle growth law, a growth law which depends on all the time scales and allocation parameters in the cycle. The relative abundance growth law is derived directly eigenvalue equation. For example, assuming that limitation regime $i$ is consistent, we know that $M_i \vec{S}_i = \mu \vec{S}_i$. Dividing by the $k^{th}$ component of $\vec{S}_i$, namely, by $(\vec{S}_i)_k$, we obtain that, for the $l^{th}$ component, $\mu = \sum_j (M_i)_{lj} \frac{(\vec{S}_i)_j}{(\vec{S}_i)_k}$. This is useful if the relative abundances $\frac{(\vec{S}_i)_j}{(\vec{S}_i)_k}$ can be measured experimentally, as in the case of the ribosomal protein mass fraction. Evidently, because all the components grow exponentially at the same rate, many more relative abundance growth laws can be derived, involving other relative abundance ratios, e.g., protein to mRNA, or RNA polymerase to mRNA. 

Before moving to explain Fig. \ref{FigMethods}B, we note that using ODEs that switch between regimes is a well-known practice in control theory under the name piecewise-smooth dynamical system or switching dynamical systems \cite{SwitchingBook}, but also in classical physics, e.g., when considering the bouncing of a basketball on the floor. In particular, the use of a minimum function is considered standard in input--output economy, where it is called Leontif's production function \cite{Leontif}. 

So far, we explained how we devise a model that couples two autocatalytic cycles, using the minimum function to define the various regimes that support balanced exponential growth. Next, we explain how we add more biological details to the model, so as to facilitate its use for deriving biologically relevant growth laws. We turn to Fig. \ref{FigMethods}B, which depicts the next step in building our model. For the sake of clarity, we simplify the model further by taking $\alpha=1$ and assuming $F \gg U$. This simplified model comprises only one autocatalytic cycle; however, we add new generic states in order to make it more realistic. First, we differentiate between two states of the $U$ machines, idling and active. The number of idling $U$'s is denoted by $U_0$. The idling state is characterized by a time scale $\tau_0$, which is the average idling duration. When $U$'s are not idling, they are active. The number of active $U$'s is denoted by $U_b$. The time scale for being active is a complex function of the allocation towards all the synthesis tasks that $U$ is allocated to, and of their respective work-loads. In this simplified picture, since we used $\alpha=1$, there is only one task that $U$ is allocated to perform---which is to replicate itself. Hence, the time scale for being active is the time scale for making new $U$'s, that is, $\tau_U = F_U \frac{\tau_{aa}}{\eta}$. 

Finally, we add a self-assembly step by further breaking the process of making new $U$'s to the process of synthesizing $a_1$, $a_2$, and $a_3$ subunits  of $U$, and to the process of assembling them to form a new $U$. This self-assembly process proceeds at a rate $\tau_{SA}^{-1}$. In our model, we use what we dub a ``Tetris'' model, which assumes that whenever a stoichiometric series of $a$ subunits, that is $a_1=a_2=a_3=1$ is formed for the first time, the assembly of a new $U$ is initiated with these subunits. The assembly will be completed, on average, $\tau_{SA}$ time units after the assembly is initiated. 

In the Results section, we combine the autocatalytic cycles of the transcription--translation machinery with other cycles, while accounting for the couplings between the different cycles, the existence of different allocation parameters for RNA polymerases and ribosomes, the existence of idling and active periods, ribosome and RNA polymerase self-assembly steps, and the finite lifetime of mRNAs, ribosomes, RNA polymerases, and proteins. The resulting model is also piecewise linear, albeit much larger, and we derive from it the various growth laws that we present above, by algebraically calculating the characteristic polynomial roots (closed-cycle growth law), or by using the eigenvector equation (relative abundance growth law) for the appropriate limitation regimes. 

\showmatmethods{} 

\acknow{We would like to thank Terry Hwa and Suckjoon Jun for critical remarks at the early stages of this work. We thank Matteo Marsili, Yinnon M. Bar-On, and Yitzhak Fishov for useful comments. This research was supported by the ISRAEL SCIENCE FOUNDATION (grant No. 776/19).}

\showacknow{} 

\bibliography{pnas-sample}

\end{document}